\documentclass[journal]{IEEEtran}
\ifCLASSINFOpdf
\else
\fi
\usepackage{graphicx} 
\usepackage{epsfig} 
\usepackage{amsmath} 
\usepackage{amssymb}  
\usepackage{psfrag}
\usepackage{color}
\usepackage{cite}
\usepackage{arydshln} 
\usepackage{multirow}

\graphicspath{{Figures/}}

\newcommand{\tA}{\mathcal{A}}
\newcommand{\tB}{\mathcal{B}}
\newcommand{\tC}{\mathcal{C}}
\newcommand{\tD}{\mathcal{D}}
\newcommand{\tR}{\mathcal{R}}
\newcommand{\tP}{\mathcal{P}}
\newcommand{\tQ}{\mathcal{Q}}
\newcommand{\tF}{\mathcal{F}}
\newcommand{\tG}{\mathcal{G}}
\newcommand{\tE}{\mathcal{E}}

\newtheorem{thm}{Theorem} 
\newtheorem{prop}{Proposition} 

\newtheorem{coro}{Corollary}

\newtheorem{rem}{Remark}

\begin{document}
%
\title{Hierarchical Decentralized Robust Optimal Design for \textcolor{blue}{Homogeneous} Linear Multi-Agent Systems }
%
%
%

\author{Dinh~Hoa~Nguyen,~\IEEEmembership{Member,~IEEE,} Tatsuo~Narikiyo,~Michihiro~Kawanishi,~\IEEEmembership{Member,~IEEE,} and~Shinji~Hara,~\IEEEmembership{Fellow,~IEEE,} 

\thanks{Dinh Hoa Nguyen, Tatsuo Narikiyo, and Michihiro Kawanishi are with Control System Laboratory, Department of Advanced Science and Technology, Toyota Technological Institute, 
        2-12-1 Hisakata, Tempaku-ku, Nagoya 468-8511, Japan. E-mail: hoadn.ac@gmail.com, n-tatsuo@toyota-ti.ac.jp, kawa@toyota-ti.ac.jp.} 
\thanks{Shinji Hara is with Department of Information Physics and Computing, Graduate School of Information Science and Technology, 
	The University of Tokyo, 7-3-1 Hongo, Bunkyo-ku, Tokyo 113-8656, Japan. E-mail: Shinji\_Hara@ipc.i.u-tokyo.ac.jp.	}			
				
} 

\newcounter{MYtempeqncnt}

\maketitle


\begin{abstract}

This paper proposes novel approaches to design hierarchical decentralized robust controllers for \textcolor{blue}{homogeneous} linear multi-agent systems (MASs) perturbed by disturbances/noise. 
Firstly, based on LQR method, we present a systematic procedure to design hierarchical decentralized optimal stabilizing controllers for MASs without disturbances/noise. 
Next, a method for deriving reduced-order hierarchical decentralized stabilizing controllers is presented by suitable selections of the weighting matrices in the LQR performance index. 
Secondly, the hierarchical decentralized robust controller designs in terms of $H_{\infty}$ and $H_{2}$ norms are introduced, which include two different scenarios namely general and LQR-based synthesis. 
For the general synthesis, the robust controller gains are computed as solutions of a distributed convex optimization problem with LMI constraints. 
On the other hand, for the LQR-based design, the robust controller gains obtained from the general synthesis are further verified as LQR stabilizing gains to be unified with the LQR-based 
design when there are no disturbances/noise. This results in a hierarchical decentralized inverse optimal control problem, for which we will propose a new method to resolve it.   
Finally, several numerical examples are presented to illustrate the effectiveness of the proposed approaches. 

\end{abstract}



\section{Introduction}
\label{intro}
%
%
%
%


 %

%
%

Multi-agent systems (MASs) have gained much attention recently since there are a lot of practical applications, e.g.,     
power grids, wireless sensor networks, transportation networks, systems biology, etc, can be formulated, analyzed and synthesized under the framework of MASs. 
A key feature in MASs is the achievement of a global objective by performing local measurement and control at each agent and simultaneously collaborating among agents using those local information.   
This allows us to design, operate, control, and manage large-scale systems in distributed and/or decentralized manners with
significantly reduced time and effort, and improved reliability and security.
 
One important class of control problems in MASs is the synthesis of distributed controllers to guarantee the MASs’
robustness to disturbances/noise which can be represented by typical system norms such as $H_{\infty}$ and $H_{2}$ norms. 
For linear MASs, there have been several studies on distributed robust $H_{\infty}$ and $H_{2}$ controllers design, e.g. \cite{Zelazo:2011,Massioni:2009,Ghadami:2013,Z.Li:2011}. 
In \cite{Zelazo:2011}, a method was presented to compute the $H_{\infty}$ norm of MASs in presence of exogenous disturbances and to synthesize the
interconnection topology, however the design of control inputs was not considered. The article \cite{Massioni:2009} introduced a synthesis
approach for $H_{\infty}$ and $H_{2}$ controllers for discrete-time MASs with decomposable structures. The design method in \cite{Massioni:2009} was
then extended in \cite{Ghadami:2013} for continuous-time MASs, which actually considered one more constraints of letting all eigenvalues
of the closed-loop MASs have real parts less than a given negative bound. The authors in \cite{Z.Li:2011} investigated the design of
$H_{\infty}$ and $H_{2}$ controllers for MASs with the same feedback matrix and coupling strength for both local feedback and cooperative
terms in the controller.

\textcolor{blue}{
On the other hand, this paper investigates the hierarchical decentralized stabilization problem for homogeneous linear MASs. 
Each agent of course can be stabilized by itself, however we consider the context where agents collaborate with others to achieve a cooperative stabilization. 
The motivations for such a cooperation are as follows.  
First, by collaborating with others, the transient responses of agents can be improved and the control performance of the whole MAS is hence better \cite{Nguyen-Hara:2015j2}. 
Second, in many practical MASs, there are coordination requirements or global constraints on the inputs, states, and outputs of agents. 
Examples include stable formation control of agents \cite{Lafferriere:2005,Fax:2004} where agents are required to form and keep a specific shape, 
and output power coordination among wind turbines in a wind farm \cite{Madjidian:2014}, just to name a few. 
Therefore, agents need to exchange information with others to achieve those global requirements or constraints that cannot be fulfilled by fully decentralized control approaches. 
}

\textcolor{blue}{
Consequently, we propose a framework to design hierarchical decentralized stabilized controller and hierarchical decentralized robust optimal $H_{\infty}$ and $H_{2}$ controllers
for homogeneous linear MASs based on LQR method. To the best of our knowledge, there has not been any similar framework in literature hitherto. 
Some previous studies, e.g. \cite{Cao:2010,Borrelli:2008,Motee:2008,Tsubakino:2013,Mosebach:2014} investigated the analysis and synthesis of linear MASs utilizing LQR approach, 
however none of them considered the robust design with respect to disturbance or noise. 
On the other hand, robust $H_{\infty}$ and $H_{2}$ controllers have also been investigated for large-scale systems, e.g. \cite{Andrea:2003,Rotkowitz:2006} but in different contexts. 
The work in \cite{Andrea:2003} considered spatially interconnected systems where the systems' structures are described by periodic or infinite interconnections, then analyzed and synthesized  
distributed $H_{\infty}$ controllers for those special systems. Next, \cite{Rotkowitz:2006} presented a general framework to characterize a broad class of optimal decentralized controllers satisfying 
a so-called quadratic invariance property that made the control design convex. However, the control design problems in this paper do not belong to the classes considered 
in \cite{Andrea:2003,Rotkowitz:2006}. 
} 

\textcolor{blue}{
The contributions of this paper are threefold.
First, this paper proposes an LQR hierarchical decentralized stabilizing control design for homogeneous linear MASs. Furthermore, 
the order of the derived controller can be reduced by appropriately choosing the LQR weighting matrices.  
Second, a new LMI-based approach is presented to design hierarchical decentralized $H_{\infty}$ and $H_{2}$ controllers for homogeneous linear MASs in presence of disturbances or noise.
The control designs are formulated as distributed convex optimization problems with LMI constraints. 
Third, the above approach is then further developed to derive hierarchical decentralized $H_{\infty}$ and $H_{2}$ controllers whose gains are optimal in LQR senses so that it is unified with 
the approach of LQR hierarchical decentralized stabilizing design.  
}
 

The following notations and symbols will be used in the paper. $\mathbf{1}_n$ represents the $n \times 1$ vector with all elements equal to $1$, and $I_n$ denotes the $n \times n$ identity matrix.  
On the other hand, the symbol $^{*}$ is utilized for denoting the complex conjugate transpose. 
Next, $\mathrm{sym}(A)$ denotes $A+A^T$ for any real matrix $A$. Moreover, $\otimes$ stands for the Kronecker product. 
Lastly, $\succ$ and $\succeq$ denote the positive definiteness and positive semi-definiteness of a matrix.

\section{Problem Description}
\label{prob-def}

\subsection{MAS Model}

Consider a perturbed MAS having $N$ identical agents whose dynamics is described by the following state space model,
\begin{equation}
	\label{agent-perturbed}
	\begin{aligned}
		\dot{x}_i &= Ax_i+Bu_i+B_dd_i, \\
		y_i &= Cx_i+Du_i, i=1,\ldots,N, 
	\end{aligned}
\end{equation} 
where $x_i \in \mathbb{R}^{n}$ is the state vector, $u_i \in \mathbb{R}^m$ is the control input, 
$y_i \in \mathbb{R}^p$ is the performance output, and $d_i \in \mathbb{R}^{n_d}$ represents the disturbance or noise in the $i$th agent, respectively; 
$A \in \mathbb{R}^{n\times n}$, $B \in \mathbb{R}^{n\times m}$, $C \in \mathbb{R}^{p\times n}$, $D \in \mathbb{R}^{p\times m}$, and $B_d \in \mathbb{R}^{n\times n_d}$. 
The perturbed MAS without controller can then be represented by
\begin{equation}
	\label{MAS-perturbed}
	\begin{aligned}
		\dot{x} &= \tA x+\tB u+\tB_dd, \\
		y &= \tC x+\tD u. 
	\end{aligned}
\end{equation} 
where $\tA=I_{N}\otimes A,\tB=I_{N}\otimes B,\tC=I_{N}\otimes C,\tD=I_{N}\otimes D,\tB_d=I_{N}\otimes B_d$, 
$x=[x_1^T,\ldots,x_N^T]^T$, $u=[u_1^T,\ldots,u_N^T]^T$, $y=[y_1^T,\ldots,y_N^T]^T$, $d=[d_1^T,\ldots,d_N^T]^T$. 
We here assume that all states of agents are measurable then our goal is to design a hierarchical decentralized state feedback controller for stabilizing the given MAS (\ref{MAS-perturbed})   
when there are no disturbances, and the effect of disturbances to the system performance when they exist, in the sense of $H_{2}$ and $H_{\infty}$ norms, is bounded by certain prescribed quantities.   


\textcolor{blue}{
As mentioned in Section \ref{intro}, the cooperation among agents is required due to some global objectives, e.g., stable formation control, output coordination, or transient response improvement, etc.  
Denote $\tG$ the graph representing the information structure required in our MAS, 
where each node in $\tG$ stands for an agent and each edge in ${\cal G}$ represents 
the interconnection between two agents. 
In this paper, we assume that 
$\tG$ is undirected. 
Then, the information structure in our MAS can be mathematically characterized by a matrix $K$ which belong to the following class,  
\begin{align}
	\label{K-class}
		\mathbb{K}_{s} \triangleq \{ & K=K^T\in\mathbb{R}^{N\times N}  \; | \; K_{ij}=0 \; \textrm{if} \; (i,j) \notin {\cal E} \; \& \; i \neq j,  \nonumber \\
		& K_{ij}=K_{ji} \neq 0 \; \textrm{if} \; (i,j) \in {\cal E} \},
\end{align} 
where $\tE$ denotes the edge set of $\tG$. }

\subsection{Hierarchical Decentralized Control Design Problems}

\begin{figure}[ht!]
		\centering
		\psfrag{u}{\Huge $u_i$}
		\psfrag{v}{\Huge $u_{i,2}$}
		\psfrag{z}{\Huge $\zeta_{i}$}
		\psfrag{zj}{\Huge $\zeta_{j}$}
		\psfrag{K}{\Huge $\displaystyle \sum_{(i,j) \in {\cal E}}{K_{ij}\zeta_{j}}$}
		\psfrag{y}{\Huge $y_i$}
		\psfrag{x}{\Huge $x_i$}
		\psfrag{d}{\Huge $d_i$}
		\psfrag{ul}{\Huge $u_{i,1}$}
		\psfrag{p}{\scalebox{1.3}{\huge Agent $i$th}}
		\psfrag{F1}{\Huge $F_{1}$}
		\psfrag{F2}{\Huge $F_{2}$}	
		\psfrag{upper}{\Huge Upper layer}
		\psfrag{lower}{\Huge Lower layer}
		\psfrag{coop}{\Huge \it \color{blue}{Cooperative control}}
		\psfrag{local}{\Huge \it \color{red}{Local control}}
		\psfrag{agg}{\Huge \it \color{green}{Aggregation}}
		\scalebox{0.32}{\includegraphics{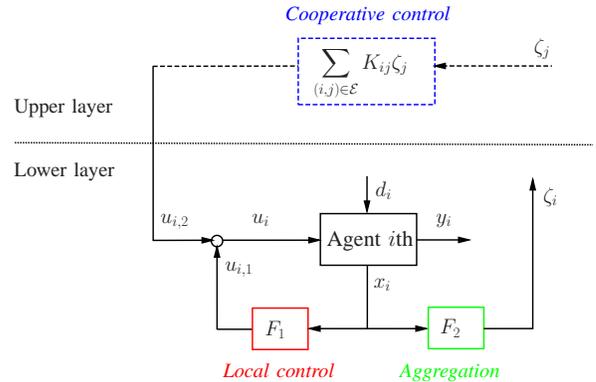}}
		\caption{Block diagram of hierarchical decentralized robust state feedback control design.}
		\label{state_fb}
	\end{figure}

Hierarchical decentralized control means that  there is the higher level interactions combined with fully decentralized  locally controlled agents.
Note that there are at least two ways to implement the higher level interactions.
One is by broadcasting the appropriate aggregated signal to the neighboring agents.
Another is by setting an authority which collects the aggregated signals from
all the agents and sending out the appropriate signals after some computation.
We prefer the former one to the latter one, because we do not need to restrict
the communication graph for the latter case.  In other words, we consider the case where MASs with a fixed
information exchange and the problem setting is on a hierarchical network structure. 
\textcolor{blue}{
In this sense, the control input to the MAS have the form
\begin{equation}
	u_i=u_{i,1}+u_{i,2}.
\end{equation} 
The first component $u_{i,1} \triangleq F_{1}x_{i} \in \mathbb{R}^{m}$ is the local feedback control from each agent to itself in order to satisfy the local objective,  
$F_{1} \in \mathbb{R}^{m \times n}$ is the local feedback gain. 
The second component $u_{i,2}\in\mathbb{R}^{m}$ is the cooperative control input obtained by exchanging information with other connected agents through its aggregated signal 
$\zeta_{i} \triangleq F_{2}x_{i} \in \mathbb{R}^{m}$ and the aggregated signal $\zeta_{j} \triangleq F_{2}x_{j} \in \mathbb{R}^{m}$ from agent $j$, $j=1,\ldots,N$, $(i,j) \in {\cal E}$, 
to achieve the global objectives in the MAS, $F_{2} \in \mathbb{R}^{m \times n}$ is the cooperative gain.  
This hierarchical decentralized state feedback control structure is illustrated in Figure~\ref{state_fb}.   }

\textcolor{blue}{
Denote $u_1 \triangleq [u_{1,1}^T, \ldots, u_{N,1}^T]^T,\zeta \triangleq [ \zeta_1^T, \ldots, \zeta_N^T]^T$, 
$u_2 \triangleq [ u_{1,2}^T,\ldots, u_{N,2}^T]^T$. 
Then the control input for the whole MAS is represented by
\begin{equation}
	\label{control-input}
	u=u_1+u_2=(I_{N} \otimes F_1)x+(K\otimes I_{m})\zeta=Fx,
\end{equation} 
where $F \triangleq I_{N} \otimes F_1 + K \otimes F_2$.  Denote $\mathbb{F}_K$ the class composing of all such structured matrix $F$, i.e., 
\begin{align}
	\label{ctlr-form}
		\mathbb{F}_K \triangleq \{ & F \in \mathbb{R}^{(Nm) \times (Nn)} \; | \; F = I_{N} \otimes F_1 + K \otimes F_2;  \nonumber \\
		& F_{1}, F_{2} \in \mathbb{R}^{m \times n}, K \in \mathbb{K}_s \}.
\end{align} 
}
Consequently, the hierarchical decentralized stabilizing control design problem is specified as follows. 

\textcolor{blue}{
\textbf{Design problem 1 (Hierarchical decentralized stabilizing controller):} 
For the given MAS (\ref{MAS-perturbed}) with $(A,B)$ controllable and without disturbances or noises, design the hierarchical decentralized controller gain $F \in \mathbb{F}_K$ such that 
the closed-loop MAS is stabilized and has a desirable information structure. This is equivalent to determine the state feedback gains $F_1, F_2 \in \mathbb{R}^{m \times n}$ 
and the cooperating gain matrix $K \in \mathbb{K}_s$.}

\textcolor{blue}{
Next, we investigate the hierarchical decentralized robust $H_{\infty}$ and $H_{2}$ controller designs to minimize the effect of disturbances or noise to the system outputs in the following problems.}

\textcolor{blue}{
\textbf{Design problem 2 (Hierarchical decentralized general robust controller):} 
For the given MAS (\ref{MAS-perturbed}) with $(A,B)$ controllable and let the cooperating gain $K \in \mathbb{K}_s$ be chosen, design the hierarchical decentralized controller gain $F \in \mathbb{F}_K$ such that it satisfies Design problem 1 when there is no disturbances/noise, and its $H_{\infty}$ or $H_{2}$ norm is upper bounded by a given positive quantity in presence of disturbances/noise.}

\textcolor{blue}{
Note that for above problem, our purpose is to determine the robust state feedback gain $F$ 
without an LQR sense. However, if the consistency of the robust controller with the stabilizing controller found by LQR approach in the design problem 1 is considered, 
$F\in\mathbb{F}_K$ should be an LQR gain. This challenging design problem is formulated as follows.}

\textcolor{blue}{
\textbf{Design problem 3 (Hierarchical decentralized LQR-based robust controller):} 
For the given MAS (\ref{MAS-perturbed}) with $(A,B)$ controllable and let the cooperating gain matrix $K \in \mathbb{K}_s$ be chosen, find a hierarchical decentralized controller gain 
$F \in \mathbb{F}_K$ which solves the design problem 2 and simultaneously is an LQR gain with respect to some performance index.}

The design problems 2 and 3 will be resolved and presented with details in Section \ref{Hinf-syn} and Section \ref{H2-syn}.

\section{Hierarchical Decentralized LQR-based Stabilizing Design for Unperturbed MASs}
\label{opt-design}

\subsection{Class of Performance Indices}

We here introduce a solution for the design problem 1 based on LQR method with the assumption that all states of agents are measurable. 
Consider the following performance index 
\begin{equation}
	\label{p-index}
	J = J_{x,\mathcal{L}}+J_{x,\mathcal{G}}+J_{u},
\end{equation} 
where
\begin{align*}
	J_{x,\mathcal{L}} &= \int_{0}^{\infty}{x^T(I_{N}\otimes Q_1)x \; dt}  & \textrm{(local performance index)},  \\
	J_{x,\mathcal{G}} &= \int_{0}^{\infty}{x^T(K\otimes Q_2)x \; dt} & \textrm{(global performance index)},  \\
	J_{u} &= \int_{0}^{\infty}{u^T\tR u \; dt}  & \textrm{(control input penalty)}, 
\end{align*}
where 
$Q_1\in\mathbb{R}^{n\times n},Q_2\in\mathbb{R}^{n\times n}$, $Q_1\succeq 0,Q_2\succeq 0$; 
$\tR\in\mathbb{R}^{(Nm)\times(Nm)}$, $\tR\succ 0$; 
$K\in\mathbb{K}_{s}^{+}$ which is a class of positive semidefinite cooperation defined as follows,
\begin{equation}
	\label{K-pos}
	\mathbb{K}_{s}^{+} \triangleq \{K \in \mathbb{K}_s \; | \;  K \succeq 0 \}.
\end{equation}
$J_{x,\mathcal{L}}$ is a local performance index composing of the individual penalties for the states of agents.  
$J_{x,\mathcal{G}}$ corresponds to a global performance index taking into account the cooperation 
among agents represented by matrix $K$.  
$J_{u}$ is a penalty for the control input required to the whole MAS. 

Next, rewriting the performance index (\ref{p-index}) as follows,
\begin{equation}
	J = \int_{0}^{\infty}{(x^T\tQ x+u^T\tR u)dt},
\end{equation}
where
\begin{equation}
	\label{w-matrices-1}
		\tQ = I_{N}\otimes Q_1+K\otimes Q_2,
\end{equation}
we aim at finding $u$ having the form (\ref{control-input}) that minimizes the performance index (\ref{p-index}). 
Employing the following assumptions:
\begin{itemize}
 \item[{\bf  A1:}]  $(\tA,\tB)$ is controllable,
 \item[{\bf  A2:}]  $(\tQ^{1/2},\tA)$ is observable,
\end{itemize}
it is shown from the optimal control theory \cite{Anderson:1990} that such an LQR controller 
is computed by $u=-Fx$, $F\in\mathbb{R}^{(Nm)\times (Nn)}$ where
$$F=\tR^{-1}\tB^T\tP,$$ with $\tP\in\mathbb{R}^{(Nn)\times (Nn)}$ is the unique positive definite solution of the following Riccati equation
\begin{equation}
	\label{Riccati-eq}
	\tP\tA+\tA^T\tP+\tQ-\tP\tB\tR^{-1}\tB^T\tP=0.
\end{equation}
Motivated by the class of distributed feedback gains (\ref{ctlr-form}) and 
the structure of the weighting matrix $\tQ$, 
we select the weighting matrix $\tR$ with the following form
\begin{equation}
	\label{w-matrices-2}
	\tR^{-1} = I_{N}\otimes R_{1}+K\otimes R_2,
\end{equation}
where $R_1,R_2 \in \mathbb{R}^{m\times m}$, $R_1 \succ 0,R_2 \succ 0$.

In the previous works, it was proved that if $\tA,\tB,\tC,\tR,\tQ$ belong to some operator algebra \cite{Motee:2008} or semigroup \cite{Tsubakino:2013} then the solution $\tP$ of the Riccati equation 
(\ref{Riccati-eq}) also belongs to that algebra or semigroup. As a result, they could prove that the LQR controller gain $\tF$ has a similar property. 
However, in our current setting, $\tA,\tB,\tC,\tR,\tQ$ do not belong to any operator algebra or semigroup, which leads to the impossibility for showing that with the choice of the weighting matrices as in (\ref{w-matrices-1}) and (\ref{w-matrices-2}), $\tP$ has the same structure. 
\textcolor{blue}{ Also, our controller class does not satisfy the quadratic invariance property in \cite{Rotkowitz:2006}.} 
Therefore, in the next subsection, we will propose a method of choosing the weighting matrices 
$\tQ$ and $\tR$ with the forms of (\ref{w-matrices-1}) and (\ref{w-matrices-2}), 
which completely fits our situation and purpose. 

\textcolor{blue}{
On one hand, our method has a disadvantage since $\tQ$ and $\tR$ must have some special forms and cannot be arbitrarily as in the standard LQR. 
Note however that this is reasonable since $\tQ$ and $\tR$ are selectable parameters, and in fact we still have some specific degree of freedom to choose them as pointed out in Step 1 and Step 3 in the design procedure below.   
On the other hand, this is also an advantage since this sacrifice on LQR cost allows us to deal with a broader class of systems where the agents' dynamics and the class of controllers 
are less restrictive than those in existing results including operator algebra, semi-group, quadratic invariance, etc. 
}

\subsection{Design Procedure}
\label{state-procedure}

In this section, we propose a systematic design procedure for state feedback hierarchical decentralized controllers as stated in the design problem 1, which consists of four steps. 
\begin{itemize}
	\item {\bf Step 1 (Local LQR Design) :} \\ 
Select the weighting matrices $Q_1 \in \mathbb{R}^{n\times n}$ and $R_1 \in \mathbb{R}^{m\times m}$ for the local objectives, 
 such that $Q_1 \succeq 0$, $R_1 \succ 0$, and $(Q_1^{1/2}, A)$ is observable,  
and solve the local Riccati equation 
\begin{equation}
	\label{R-eq-local}
	P_1A+A^TP_1-P_1BR_1B^TP_1+Q_1=0, 
\end{equation}
to obtain the unique positive definite solution $P_1 \in \mathbb{R}^{n\times n}$. 

  \item  {\bf Step 2 (Setting Upper Layer Interactions) :} \\ 
Choose a positive semidefinite matrix $K\in\mathbb{K}^{+}_{s}$.

	\item {\bf Step 3 (Global LQR Setting) :} \\ 
	Choose $R_2 \in \mathbb{R}^{m\times m}$, $R_2 \succ 0$ and set $Q_2\in\mathbb{R}^{n\times n}$ as follows:
	\begin{equation}
		\label{Q2}
		Q_2 = P_1BR_2B^TP_1,
	\end{equation}
	where $P_1\in\mathbb{R}^{n\times n}$ is obtained from Step 1. 
	
	\item  {\bf Step 4 (State Feedback Gain Calculation) :} \\ 
Set the state feedback gains $F_1$ and $F_2$ as follows: 
\begin{equation*}
		F_1 = R_1B^TP_1, ~
		F_2 = R_2B^TP_1.
\end{equation*}
\end{itemize}

\textcolor{blue}{
\begin{rem}
Note that in the Step 2 above, even when the MAS's information structure is given, i.e., the structure of $K$ is known, we still can design $K$ by choosing its elements. 
This will be illustrated in the numerical example in Section \ref{num-1}. 
\end{rem}
}

The following theorem clearly shows the validation of the above procedure in which
the resultant LQR controller belong to the class $\mathbb{F}_K$ in (\ref{ctlr-form})  
if the weighting matrices are chosen as in the design procedure. 

\begin{thm}
\label{hier-ctlr}
Consider the MAS with dynamics of agents represented by (\ref{agent-perturbed}) 
in which $(A,B)$ is controllable. 
Let $K$ be a matrix in the class $\mathbb{K}_{s}^{+}$ and the weighting matrices $\tQ$ and $\tR$ 
have the forms (\ref{w-matrices-1}) and (\ref{w-matrices-2}) 
with $R_2 \in \mathbb{R}^{m\times m}$, $R_2 \succ 0$ and $Q_2\in\mathbb{R}^{n\times n}$ chosen as in (\ref{Q2}). 
Then the hierarchical decentralized optimal LQR state feedback gain is given by
\begin{equation}
	\label{lqr-controller}
	F=I_{N}\otimes F_1+K\otimes F_2,
\end{equation}
where
\begin{equation}
	\label{gains}
		F_1 = R_1B^TP_1,
		F_2 = R_2B^TP_1.
\end{equation}
\end{thm}


\begin{IEEEproof}
Assumption {\bf A1} is obvious since it is equivalent to the controllability of $(A,B)$. 
On the other hand, Assumption {\bf A2} holds when $Q_2 = 0$. With extra term $K \otimes Q_2$ in $\tQ$, which is positive semidefinite, the observability condition is not broken,  
and hence Assumption {\bf A2} holds even for any $Q_2 \succeq  0$. 
In addition, $\tR^{-1} = I_{N}\otimes R_1+K\otimes R_2 \succ 0$, i.e., $\tR\succ 0$.
Thus, (\ref{Riccati-eq}) has a unique positive definite solution.

Next, substituting $\tP=I_{N}\otimes P_1$ and $\tQ,\tR$ back to the Riccati equation (\ref{Riccati-eq}), we obtain 
\begin{align}
	\label{R-1}
	0 = & \, I_{N}\otimes(P_1A+A^TP_1-P_1BR_1B^TP_1+Q_1) \nonumber \\
	& +K\otimes(Q_2-P_1BR_2B^TP_1).  
\end{align} 
This is always true with $Q_2=P_1BR_2B^TP_1$ and $P_1$ is the solution
of (\ref{R-eq-local}). Hence, $\tP=I_{N}\otimes P_1$ is a solution of (\ref{Riccati-eq}). 
Since we have obtained the uniqueness of the solution of (\ref{Riccati-eq}), that matrix $\tP$ is the unique solution. Accordingly, the hierarchical decentralized optimal LQR controller gain is calculated as follows,
\begin{align*}
	F &= \tR^{-1}\tB^T\tP, \\
	&= (I_{N}\otimes R_1+K\otimes R_2)(I_{N}\otimes B^T)(I_{N}\otimes P_1), \\
	&= I_{N}\otimes(R_1B^TP_1)+K\otimes(R_2B^TP_1).
\end{align*} 
Thus, $F_1$ and $F_2$ are determined by (\ref{gains}).
\end{IEEEproof}


The main advantages of the proposed controller in Theorem \ref{hier-ctlr} are twofold. Firstly, it is hierarchical decentralized, which composes of two components. The first component with the gain 
$I_{N}\otimes(R_1B^TP_1)$ represents the local feedback of each agent to itself in order to satisfy the local objective of its stability. 
\textcolor{blue}{The second component with the gain $K\otimes(R_2B^TP_1)$ shows the cooperation among agents to achieve the global objectives of the whole MAS (e.g. in stable formation control with a desired information structure). 
More specifically, matrix $K$, which is selectable as in Step 2 of the design procedure in Section \ref{state-procedure}, reflects the communication structure among agents; and the other term $R_2B^TP_1$ 
shows the form of aggregated signals to be exchanged among agents.} 
Secondly, the proposed controller is globally optimal but its gains are locally calculated by solving just a common local low-dimension Riccati equation (\ref{R-eq-local}), which is independent of the number of agents. Therefore, the computation burden as the number of agents largely increases is completely avoided. 

On the other hand, it can be seen that the optimal performance index for the whole MAS is equal to 
\begin{equation}
	{\cal J}_{\min} = x_0^TPx_0 = \sum_{i=1}^{N}x_{0,i}^TP_1x_{0,i},
\end{equation} 
which is sum of minimum local performance indices of individual agents and is independent of the information graph. Therefore, the cost for stabilization is not changed even if the communication structure in the MAS is varied. 

%

\begin{rem}
Another result of LQR-based hierarchical decentralized stabilizing controllers for heterogeneous MASs has recently been introduced in \cite{Nguyen-Hara:2015j2} of which agents have the same input dimension but the state and output dimensions are different. 
However, the derived controller in \cite{Nguyen-Hara:2015j2} restricted the class of the LQR weighting matrices, which resulted in a restricted hierarchical decentralized controller 
(cf. Theorem 1 in \cite{Nguyen-Hara:2015j2}).  
More specifically, when applying to homogeneous MASs, the controller in \cite{Nguyen-Hara:2015j2} becomes less general than the obtained one (\ref{lqr-controller}) in the current paper, where  
$R_1$ and $R_2$ become multiples of the identity matrix. 
Furthermore, the designs of reduced-order hierarchical decentralized controllers, robust $H_{\infty}$ and $H_{2}$ hierarchical decentralized controllers 
were not considered in \cite{Nguyen-Hara:2015j2}. 
\end{rem}

\subsection{\textcolor{blue}{Illustrative example 1} }
\label{num-1}

\textcolor{blue}{
Consider a network of $3$ identical unstable agents whose model is described by (\ref{agent-perturbed}) without disturbances or noise and  
\begin{equation}
	A=\begin{bmatrix} 0 & 1 \\ -1 & -1 \end{bmatrix},B=\begin{bmatrix} 0 \\  1 \end{bmatrix}.
\end{equation}
The outputs of agents are $y_{i}=Cx_{i},i=1,2,3$ where $C=[1,0]$. 
Then let $K$ be a Laplacian matrix as follows,
\begin{equation}
	\label{c3-eq1}
	K=\begin{bmatrix} 1 & -1 & 0 \\ -1 & 1+q & -q \\ 0 & -q & q \end{bmatrix}, q\geq 0.
\end{equation}
This matrix $K$ implies that agents $1$ and $2$ are connected, agents $2$ and $3$ may be connected while agents $1$ and $3$ are not connected. 
Subsequently, we can rewrite the global performance index as follows,
\begin{equation*}
	J_{x,\mathcal{G}}=(x_1-x_2)^TQ_{2}(x_1-x_2)+q(x_2-x_3)^TQ_{2}(x_2-x_3).
\end{equation*}
It then can be seen that $J_{x,\mathcal{G}}$ puts a penalty under the form of a quadratic function with a weight matrix $Q_{2}$ for the differences between the states of agents and hence by minimizing $J$ including $J_{x,\mathcal{G}}$, the gaps between agents' states are reduced simultaneously with the decrease of the states. As a result, the convergence speed of the agents' states to zero will be faster as both $J_{x,\mathcal{L}}$ and $J_{x,\mathcal{G}}$ are utilized than when only $J_{x,\mathcal{L}}$ is used. Furthermore, by changing the value of $q$, the system responses are also changed. 
}

\begin{figure}[ht!]
		\centering
		\includegraphics[scale=0.45]{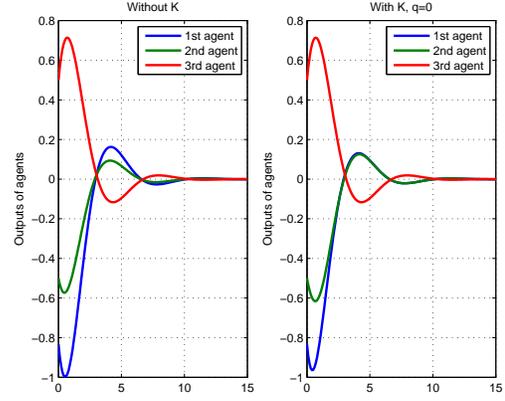}
		\caption{\textcolor{blue}{System responses without (left) and with (right) a global performance index but $q=0$.} }
		\label{c3_eg_1}
	\end{figure}

\begin{figure}[ht!]
		\centering
		\includegraphics[scale=0.45]{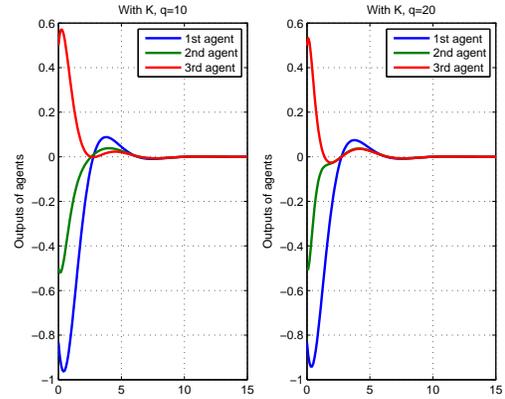}
		\caption{\textcolor{blue}{System responses with a global performance index as $q=10$ (left) and $q=20$ (right).} }
		\label{c3_eg_2}
	\end{figure}

\textcolor{blue}{
The simulation result in Figure \ref{c3_eg_1} exhibits the output responses of agents without a global performance index and when a global performance index is employed with $q=0$, i.e., agents $2$ 
and $3$ are not connected. Obviously, when the global performance index is employed but $q=0$, the output of agent $1$ converges to the output of agent $2$ before all outputs of agents come to zero. 
This is because only those agents are connected while agent $3$ is not connected to any of them. Next, Figure \ref{c3_eg_2} shows the output responses of agents when $q=10$ and $q=20$. 
We can observe that the convergence speed in these two cases are faster than in the former ones. Furthermore, the output of agent $2$ rapidly converges to the output of agent $3$ before all the agents' outputs come to zero as $q$ increases. This is because a much larger weight is put on the state difference of the $2$nd and $3$rd agents making them converge to each other faster. 
In other words, by letting $q$ larger the network is divided into two groups of agents in which the first group consists of agent $1$ and the second group composes of agents $2$ and $3$. 
{\it Thus, the structure of the network is clearly reflected in the interconnection matrix $K$ and changing the elements of $K$ can improve the transient responses of agents and the whole MAS's control performance.}}

\subsection{ \textcolor{blue}{Further Stabilizing Design: Reduced-order Hierarchical Decentralized Controllers} }
\label{reduced-design}

In this section, we aim at developing a method to reduce the order of the hierarchical decentralized optimal stabilizing controller derived above.  
Following the idea of selective pole shift proposed in \cite{Kawasaki:1983,Kawasaki:1989,Kraus:1999,Cigler:2009} for single linear systems, we further elaborate 
on the suitable selection of the LQR weighting matrices to derive the reduced-order hierarchical decentralized stabilizing controllers for linear MASs. 
The details of our method is presented below. 


For simplicity, we assume that the eigenvalues of each agent's state matrix $A$ are distinct. 
Suppose that $(\lambda_{1},\ldots,\lambda_{q})$, $0<q<n$ are unstable or undesirable eigenvalues of $A$ and 
$\nu_{1}^{*},\ldots,\nu_{q}^{*}$ are the associated left-eigenvectors. 
Let us select the weighting matrix $Q_{1}$ under the form
\begin{equation}
	\label{Q1}
	Q_{1}=V \tQ_{1} V^{*},
\end{equation}  
where $V=\begin{bmatrix} \nu_{1} & \cdots & \nu_{q} \end{bmatrix}$, $\tQ_{1} \succeq 0$. 
We will show in the following theorem that this selection of $Q_1$ gives us a 
reduced-order hierarchical decentralized controller for the MAS (\ref{MAS-perturbed}) without disturbances or noise.

\begin{thm}
\label{eigens-hetero}
With $Q_{1}$ selected in (\ref{Q1}) such that assumption A2 is satisfied, the controller designed in Theorem \ref{hier-ctlr} has the order of at most $q<n$, resulting in a reduced-order hierarchical decentralized optimal LQR state feedback controller for the given MAS (\ref{MAS-perturbed}) without disturbances or noise. Moreover, the eigenvalue set the closed-loop system matrix 
$\mathbb{A}$ is characterized as follows,
\begin{equation}
	\label{eigen-set-A}
	\sigma(\mathbb{A})=
	\left(\bigcup_{\gamma \in \sigma(K)}\sigma(\Xi_{\gamma})\right)\bigcup\left(\sigma(A)\backslash \{\lambda_{1},\ldots,\lambda_{q}\}\right),
\end{equation}
where $\Xi_{\gamma}$ is defined by
\begin{equation}
	\label{xi-hetero}
	\Xi_{\gamma}=\Lambda- z(R_1+\gamma R_2)z^{*}\tP_{1}, \Lambda= \mathrm{diag}\{\lambda_{i}\}_{i=1,\ldots,q},
\end{equation}
with $z^{*} \triangleq B^T V \in\mathbb{C}^{q\times m}$. 
\end{thm}

\begin{IEEEproof}
Let $\tP_{1}\in\mathbb{R}^{q\times q}, \tP_{1} \succ 0$ be the solution of the following Riccati equation   
\begin{equation}
	\label{R-eq-3}
	\tP_{1}\Lambda+\Lambda\tP_{1}-\tP_{1}\tR_{1}\tP_{1}+\tQ_{1}=0,
\end{equation}
where $\tR_{1} \triangleq V^{*}BR_{1}B^TV.$ 
Then we can easily check that $P_{1}=V\tP_{1} V^{*}, P_1\succ 0$ is a solution of the Riccati equation (\ref{R-eq-local}) by substituting $Q_{1}$ and $P_{1}$ back to (\ref{R-eq-local}). 
Since Riccati equation (\ref{R-eq-local}) has a unique positive definite solution, this $P_1$ is indeed that unique solution. Hence, instead of solving an $n$th-order Riccati equation (\ref{R-eq-local}), 
we only need to resolve a lower dimension, $q$th-order Riccati equation (\ref{R-eq-3}). As a result, the controller designed in Theorem \ref{hier-ctlr} becomes a reduced-order hierarchical decentralized optimal LQR state feedback controller for the given MAS (\ref{MAS-perturbed}) without disturbances or noise.  

Then the eigenvalue spectrum of the closed-loop system matrix $\mathbb{A}$ can be determined as follows. Suppose that $\eta_j\in\mathbb{C}^{n}$ is any right eigenvector of $A$ corresponding to an eigenvalue $\lambda_j\in\sigma(A)\backslash\{\lambda_{1},\ldots,\lambda_{q}\}$, $j=q+1,\ldots,n$. 
For $i=1,\ldots,N$, denote $e_i\in\mathbb{R}^{N}$ the vector whose $i$th element is $1$ while all other elements are zero. Then
\begin{align}
	\label{eig-eq1}
	\mathbb{A}(e_i \otimes \eta_j) &= e_i\otimes[(A-BR_1B^TP_{1})\eta_j)] \nonumber \\
	& \quad -(Ke_i)\otimes(BR_2B^TP_{1}\eta_j), \nonumber \\
	&= \lambda_j(e_i\otimes \eta_j)-e_i\otimes(BR_1B^TV\tP_{1}V^{*}\eta_j) \nonumber \\
	& \quad -(Ke_i)\otimes (BR_1B^TV\tP_{1}V^{*}\eta_j), \nonumber \\
	&= \lambda_j(e_i\otimes \eta_j),
\end{align}
due to a fact that $V^{*}\eta_j=0$ since $\nu_{l}^{*}\eta_j=0 \;\forall\; l=1,\ldots,q.$ 
Therefore, $\lambda_j$ is an eigenvalue of $\tA$ with the associated right eigenvector $e_i\otimes \eta_j$. This leads to
\begin{equation}
	\label{eig-eq2}
	\left(\sigma(A)\backslash \{\lambda_{1},\ldots,\lambda_{q}\}\right) \subset \sigma(\mathbb{A}),
\end{equation} 
with the note that each eigenvalue $\lambda_j,j=q+1,\ldots,n$ has the multiplicity $N$. 
On the other hand, 
\begin{align*}
	& (I_N\otimes V^{*})\mathbb{A} \nonumber \\
	&= (I_N\otimes V^{*})[I_{N}\otimes (A-BR_1B^TP_1)-K\otimes(BR_2B^TP_1)] \nonumber \\
	&= I_N\otimes (\Lambda V^{*}-zR_1z^{*}\tP_{1} V^{*})-K\otimes(zR_2z^{*}\tP_{1} V^{*}), \nonumber \\
	&= [I_N\otimes (\Lambda-zR_1z^{*}\tP_{1})-K\otimes(zR_2z^{*}\tP_{1})](I_N\otimes V^{*}). 
\end{align*}
Let $\rho^{*}$ is any left eigenvector of $K$ and $\gamma$ is the associated eigenvalue. 
Then multiplying both sides of last equation above with $\rho^{*}\otimes \xi^{*}$, $\xi \in \mathbb{C}^{q}$, we obtain
{\small
\begin{align*}
	& (\rho^{*}\otimes \xi^{*})(I_N\otimes V^{*})\mathbb{A}, \\ 
	&= [\rho^{*}\otimes(\xi^{*}\Lambda-\xi^{*}zR_1z^{*}\tP_{1})-\gamma\rho^{*}\otimes(\xi^{*}zR_2z^{*}\tP_{1})](I_N\otimes V^{*}), \\
	&= \rho^{*}\otimes(\xi^{*}\Xi_{\gamma})(I_N\otimes V^{*}),
\end{align*} }
where $\Xi_{\gamma}$ is defined in (\ref{xi-hetero}). Now, letting $\xi^{*}$ be a left eigenvector of $\Xi_{\gamma}$ with associated eigenvalue $\alpha$ gives us
\begin{align}
	\label{eig-eq5}
	(\rho^{*}\otimes \xi^{*})(I_N\otimes V^{*})\mathbb{A} &= \alpha(\rho^{*}\otimes\xi^{*})(I_N\otimes V^{*}),  \nonumber \\
	&= \alpha\rho^{*}\otimes(\xi^{*}V^{*}).
\end{align}
This means that $\alpha$ is also an eigenvalue of $\mathbb{A}$ with the associated left eigenvector $\rho^{*}\otimes(\xi^{*}V^{*})$. Furthermore, $\alpha$ has the multiplicity $N$. Accordingly,
\begin{equation}
	\label{eig-eq6}
	\sigma(\Xi_{\gamma}) \subset \sigma(\mathbb{A}).
\end{equation}
Thus, combining (\ref{eig-eq2}) and (\ref{eig-eq6}) gives us (\ref{eigen-set-A}).
\end{IEEEproof}

If the matrix $A$ has only one unstable eigenvalue $\lambda$, we can design a hierarchical decentralized optimal controller with order $1$ instead of a full-order hierarchical decentralized optimal controller for the MAS. This indeed gives a superior simplicity for hierarchical decentralized controllers design for large-scale MASs. Furthermore, we explicitly know all eigenvalues of the closed-loop system matrix as in Corollary \ref{coro-1}, which provides a strong basic for better evaluating the control performance as well as adjusting the design parameters to achieve an expected control performance. 

\begin{coro}
\label{coro-1}
The MAS (\ref{MAS-perturbed}) without disturbances is cooperatively stabilized by a $1$-order hierarchical decentralized optimal state feedback controller $u=-Fx$ with 
\begin{equation}
	\scalebox{0.9}{$
	F=\frac{\lambda+\sqrt{\lambda^2+q_1r_1}}{r_1}\left(I_N\otimes(R_1B^Tvv^T)+K\otimes(R_2B^Tvv^T)\right). $}
\end{equation} 
Accordingly, the eigenvalues of the closed-loop system matrix are 
\begin{equation}
	\scalebox{0.8}{$
	\displaystyle \bigcup_{\gamma \in \sigma(K)}\left(-\sqrt{\lambda^2+q_1r_1}-\gamma\frac{r_2(\lambda+\sqrt{\lambda^2+q_1r_1})}{r_1}\right) 
	\bigcup\left(\sigma(A)\backslash \{\lambda\}\right),
	$}
\end{equation}
where $r_1=v^TBR_1B^Tv,r_2=v^TBR_2B^Tv$, $v^T$ is the left eigenvector of $A$ associated with $\lambda$; $Q_1=vq_1v^T, q_1 \geq 0.$ 
\end{coro}

\section{Hierarchical Decentralized $H_{\infty}$ CONTROL}
\label{Hinf-syn}

In this section, the main focus is to design hierarchical decentralized controllers for the given MAS (\ref{MAS-perturbed}), which is robust with respect to the presence of the disturbance/noise vector 
$d$ in the sense of $H_{\infty}$ and $H_{2}$ norms, i.e., $\|G_{yd}(s)\|_{\infty}$ and $\|G_{yd}(s)\|_{2}$ are bounded by some given values, where $G_{yd}(s)$ is the transfer function of the closed-loop system from the disturbance $d$ to the performance output $y$.  

Substituting the distributed optimal stabilizing controller $u=-Fx$ where $F$ belongs to the set (\ref{ctlr-form}) into the MAS model (\ref{MAS-perturbed}) gives us the following closed-loop dynamics 
\begin{equation}
	G_{yd}(s)=\tilde{\tC}[sI_{Nn}-\tilde{\tA}]^{-1}\tB_d,
\end{equation} 
where 
\begin{align*}
	\tilde{\tC} &= I_{N}\otimes(C-DF_1)-K\otimes(DF_2), \\
	\tilde{\tA} &= I_{N}\otimes(A-BF_1)-K\otimes(BF_2).
\end{align*}
Since $K$ is a positive semi-definite matrix, there exists an orthogonal matrix $U\in\mathbb{R}^{N\times N}$ such that $K=U^{-1}\Gamma U$, where $\Gamma=\mathrm{diag}\{\gamma_1,\ldots,\gamma_N\}$ is a diagonal matrix whose diagonal elements are eigenvalues of $K$. 
Consequently, denote 
$$\tilde{x}=(U\otimes I_n)x,\tilde{y}=(U\otimes I_n)y,\tilde{d}=(U\otimes I_n)d,$$
and multiply both sides of the perturbed closed-loop model of MAS with $U\otimes I_n$, we obtain
\begin{equation}
	\label{MAS-perturbed-1}
	\begin{aligned}
		\dot{\tilde{x}} =& [I_N\otimes(A-BF_1)-\Gamma\otimes(BF_2)]\tilde{x}+\tB_d\tilde{d}, \\
		\tilde{y} =& [I_N\otimes(C-DF_1)-\Gamma\otimes(DF_2)]\tilde{x}. 
	\end{aligned}
\end{equation} 
It can be easily observed that (\ref{MAS-perturbed-1}) is in block-diagonal form which means that the perturbed closed-loop model of MAS is decomposed into $N$ independent subsystems
\begin{equation}
	\label{MAS-perturbed-2}
	\begin{aligned}
		\dot{\tilde{x}}_i =& [A-B(F_1+\gamma_iF_2)]\tilde{x}_i+B_d\tilde{d}_i, \\
		\tilde{y}_i =& [C-D(F_1+\gamma_iF_2)]\tilde{x}_i, i=1,\ldots,N.
	\end{aligned}
\end{equation}
Then it can be shown that \cite{Massioni:2009}
\begin{equation}
	\label{distributed-norms}
	\begin{aligned}
		\|G_{yd}(s)\|_{\infty} &=\|G_{\tilde{y}\tilde{d}}(s)\|_{\infty}=\max_{i=1,\ldots,N}\|G_{\tilde{y}_i\tilde{d}_i}(s)\|_{\infty}, \\
		\|G_{yd}(s)\|_{2}^{2} &=\|G_{\tilde{y}\tilde{d}}(s)\|_{2}^{2}=\sum_{i=1}^{N}\|G_{\tilde{y}_i\tilde{d}_i}(s)\|_{2}^{2}.
	\end{aligned}
\end{equation}
where $G_{\tilde{y}\tilde{d}}(s)$ is the transfer function of the transferred closed-loop system (\ref{MAS-perturbed-1}), and $G_{\tilde{y}_i\tilde{d}_i}(s)$ is the transfer function from $\tilde{d}_i$ to $\tilde{y}_i$. 

There are two points to be emphasized here. First, even though the formulas (\ref{distributed-norms}) from \cite{Massioni:2009} is adapted, the problem considered in \cite{Massioni:2009} is different from ours. Second, by using the transformation $U\otimes I_n$, we obtain a new decomposed MAS from the closed-loop MAS equipped with the designed hierarchical decentralized stabilizing controller 
(\ref{lqr-controller}) in which the new agents have same dimensions as the original agents.

Based on (\ref{distributed-norms}), the robust controller design for the MAS (\ref{MAS-perturbed}) now can be processed in a hierarchical decentralized manner, by designing the local robust controllers for new agents.  
This will be presented in details in the next section.

\subsection{General Hierarchical Decentralized $H_{\infty}$ Controller}

As seen above, the perturbed MAS (\ref{MAS-perturbed}) controlled by the hierarchical decentralized stabilizing controller 
$u=-(I_{N}\otimes F_1+K\otimes F_2)x$ has the $H_{\infty}$ norm less than a provided bound $\epsilon>0$ if and only if the new agents in (\ref{MAS-perturbed-2})
have $H_{\infty}$ norms less than $\epsilon$.  
However, to the best of our knowledge, there has been no necessary and sufficient LMI-based design of distributed or hierarchical decentralized $H_{\infty}$ or $H_{2}$ controllers for linear MASs. 
Only sufficient conditions for distributed $H_{\infty}$ and $H_{2}$ controller designs have been derived hitherto, which are all conservative, e.g. \cite{Massioni:2009,Ghadami:2013,Z.Li:2011}. 
Dilated LMI designs of distributed $H_{\infty}$ and $H_{2}$ controllers were introduced in \cite{Massioni:2009,Ghadami:2013} for discrete-time and continuous-time MASs, which tried to reduce the conservatism by using non-common Lyapunov variables but other common variables were needed. 
On the other hand, we propose in the following a sufficient condition for synthesis of a hierarchical decentralized $H_{\infty}$ controller for the perturbed MAS (\ref{MAS-perturbed}) which is formulated as an LMI problem with non-common Lyapunov variables and no additional common variable is needed. 

\begin{thm}
\label{Hinf-design-1}
Suppose that $K\in\mathbb{K}^{+}_{s}$ is chosen. Then the controller gains $F_1,F_2\in\mathbb{R}^{m\times n}$, which let the closed-loop MAS without disturbances stabilized and its $H_{\infty}$ norm  in presence of disturbance/noise less than $\epsilon$, is calculated by 
\begin{equation}	
	\label{F1-F2-Hinf}
		F_1 = G_1Y^{-1}, ~
		F_2 = G_2Y^{-1},
\end{equation}
if there exist $Y=Y^T\in\mathbb{R}^{n\times n},Y \succ 0$ and $G_1,G_2\in\mathbb{R}^{m\times n}$ satisfy the following set of LMIs,
\begin{equation}
	\label{Hinf-check}
	\begin{bmatrix}
		\mathrm{sym}(\alpha_{i}AY-\alpha_{i}B\tilde{G}_i)+\frac{1}{\epsilon^2}B_dB_d^T & \alpha_{i}(CY-D\tilde{G}_i)^T \\
		\alpha_{i}(CY-D\tilde{G}_i) & -I_{p} \\
	\end{bmatrix} \prec 0,  
\end{equation}
for all $i=1,\ldots,N$, where $\tilde{G}_i=G_1+\gamma_{i}G_2$, and $\gamma_{i}$ are eigenvalues of $K$, $\alpha_{i}$ are chosen positive parameters.	
\end{thm}

\begin{IEEEproof}
See Appendix \ref{apdx-1}. 
\end{IEEEproof}

Note here that the Lyapunov variables in the LMI problems (\ref{Hinf-check}) are $Y_{i} \triangleq \alpha_{i}Y$, $i=1,\ldots,N$. Hence, instead of using a common Lyapunov variable $Y$, we scale it to obtain different ones, which makes the design less conservative. Furthermore, the parameters $\alpha_{i}$ have significant effects to the obtained hierarchical decentralized $H_{\infty}$ controller, as will be shown in the illustrative example 2.

\subsection{LQR-based Hierarchical Decentralized $H_{\infty}$ Controller}

\begin{figure*}[!t]
\normalsize
\setcounter{MYtempeqncnt}{\value{equation}}
\setcounter{equation}{39}
\begin{equation}
	\label{dis-Hinf-ctlr-form}
		\begin{aligned}
			& 
			\begin{bmatrix}
				\mathrm{sym}(\alpha_{i}AY-\alpha_{i}B(R_1+\gamma_iR_2)B^T)+\frac{1}{\epsilon^2}B_dB_d^T & \alpha_{i}(CY-D(R_1+\gamma_iR_2)B^T)^T \\
					\alpha_{i}(CY-D(R_1+\gamma_iR_2)B^T) & -I_{p} \\
			\end{bmatrix} \prec 0,  \forall~ i=1,\ldots,N, \\	
			&  
				\begin{bmatrix}
					\mathrm{sym}(AY-BR_1B^T) & BR_1 \\
					R_1B^T & -R_1
				\end{bmatrix} \preceq 0,	
		\end{aligned}	
\end{equation}
\setcounter{equation}{\value{MYtempeqncnt}}
\hrulefill
\vspace*{4pt}
\end{figure*}

In this section, we would like to verify that the controller gains $F_1,F_2$ obtained from Theorem \ref{Hinf-design-1} could be derived from solution of any local Riccati equation (\ref{R-eq-local}) with proper weighting matrices $Q \in \mathbb{R}^{n\times n}$, $R \in \mathbb{R}^{m\times m}$ and $Q\succeq 0$, $R\succ 0$. In other words, we attempt to propose an approach to design hierarchical decentralized robust $H_{\infty}$ controllers for the perturbed MAS (\ref{MAS-perturbed}), which is unified with the proposed LQR-based approaches in Section \ref{opt-design} in the scenario of no disturbances or noise. To do so, we need to determine whether $F_1$ is a local LQR stabilizing controller gain. 
This is indeed an inverse optimal control problem for single LTI systems, which is a very classical but not many results have been obtained hitherto. Recently, this problem has gained a significant interest with applications to other control problems, for instance  \cite{Kong:2012}, \cite{Priess:2015}. The article \cite{Kong:2012} proposed a necessary and sufficient condition for the existence of a solution to the inverse LQR optimal control problem in the frequency domain with one weighting matrix set to be identity matrix. On the other hand, the authors in \cite{Priess:2015} presented a method to find the solution under the form of LMIs and affine matrix equalities, with application to biological systems analysis.

In this paper, we introduce a new formulation to find the solution to the inverse LQR optimal control problem, which is provided in the following proposition. 

\begin{prop}
\label{inverse-LQR}
For any stabilizing controller gain $\tF\in \mathbb{R}^{m\times n}$ for the agent model (\ref{agent-perturbed}) with no disturbance and $D=0$, let $Z=Z^T \in \mathbb{R}^{n\times n}$ and 
$R \in \mathbb{R}^{m\times m}$ be the solutions of the following feasible problem
\begin{equation}
	\label{Hinf-ZR-check}
	\begin{aligned}
		\exists&  \quad Z\succ 0, R\succ 0, \\
		\textrm{s.t.} & \quad 
			\begin{bmatrix}
				\mathrm{sym}(AZ-B\tF Z) & Z\tF^T \\
				\tF Z & -R
			\end{bmatrix} \preceq 0, \\
		& \quad \tF Z=RB^T.		
	\end{aligned}
\end{equation}	
Then $\tF$ is an LQR optimal stabilizing controller gain with the associated weighting matrices $R^{-1}$ and $Q=-A^TZ^{-1}-Z^{-1}A+Z^{-1}BRB^TZ^{-1}$. 
\end{prop}

\begin{IEEEproof}
Suppose that $\tF$ is the solution of an inverse LQR optimal control problem with the associated weighting matrices $Q \in \mathbb{R}^{n\times n}$ and $R^{-1} \in \mathbb{R}^{m\times m}$, 
$Q\succeq 0$, $R\succ 0$ and the associated Riccati equation
\begin{equation}
	\label{Req-revisit}
	A^TP+PA-PBRB^TP+Q=0.
\end{equation}
Then $\tF=RB^TP$. Now, let $Z=P^{-1}$ and multiply both to the left and to the right of (\ref{Req-revisit}) with $Z$, we obtain
\begin{align}
	\label{Req-revisit-1}
	& ZA^T+AZ-BRB^T+ZQZ=0, \nonumber \\
	\Leftrightarrow~ & Z(A-B\tF)^T+(A-B\tF)Z \nonumber \\
	& +Z\tF^TR^{-1}\tF Z+ZQZ=0.
\end{align}
Since $Q\succeq 0$, we can deduce from (\ref{Req-revisit-1}) that $Z(A-B\tF)^T+(A-B\tF)Z+Z\tF^TR^{-1}\tF Z \preceq 0$, which can be reformulated as the following LMI using the Schur complement 
\cite{Boyd:2004},

\addtocounter{equation}{1}

\begin{equation}
	\begin{bmatrix}
		\mathrm{sym}(AZ-B\tF Z) & Z\tF^T \\
		\tF Z & -R
	\end{bmatrix} \preceq 0.
\end{equation}
Therefore, $Z$ and $R$ are the solutions of the feasible problem (\ref{Hinf-check}). Once they are determine, we can compute $Q$ from the Riccati equation (\ref{Req-revisit-1}) by 
$Q=-A^TZ^{-1}+Z^{-1}A+Z^{-1}BRB^TZ^{-1}$. 
\end{IEEEproof}

Note here that though the optimization problems (\ref{Hinf-check}) and (\ref{Hinf-ZR-check}) can be resolved independently, we can combine those feasibility problems into a unique one, as in the following theorem, by further noting that the roles of $Y$ and $Z$, $G$ and $RB^T$ in those problems are similar. 
 
\begin{thm}
\label{Hinf-design-g}
For a given $\epsilon>0$, an interconnection matrix $K\in\mathbb{K}^{+}_{s}$, and $\alpha_{i}>0$, $i=1,\ldots,N$, let $Y=Y^T\in\mathbb{R}^{n\times n},Y\succ 0$ and 
$R_1,R_2 \in \mathbb{R}^{m\times m},R_1=R_1^T,R_2=R_2^T,R_1,R_2\succ 0$ be the solutions of the set of LMIs (\ref{dis-Hinf-ctlr-form}), 
where $\gamma_i,i=1,\ldots,N,$ are eigenvalues of $K$. 
Then the distributed controller $u=-[I_{N}\otimes(R_1B^TY^{-1})+K\otimes(R_2B^TY^{-1})]$ makes the MAS (\ref{MAS-perturbed}) without disturbances stabilized and its $H_{\infty}$ norm in presence of disturbances less than $\epsilon$. Furthermore, this distributed controller is globally optimal with respect to the LQR performance index (\ref{p-index}) with the weighting matrices defined in 
(\ref{w-matrices-1}) and (\ref{w-matrices-2}), where $Q_1=-A^TY^{-1}-Y^{-1}A+Y^{-1}BR_{1}B^TY^{-1}$ and $Q_2=Y^{-1}BR_2B^TY^{-1}$. 
\end{thm}

It should be emphasized that the feasibility problems in Theorem \ref{Hinf-design-1} and Proposition \ref{inverse-LQR} can have multiple solutions, and hence so is Theorem \ref{Hinf-design-g}, i.e., there may exist many hierarchical decentralized robust $H_{\infty}$ controllers.  
Thus, in order to derive a unique one, we propose the following convex optimization problem with additional slack variables \cite{Boyd:2004} for minimizing the condition numbers of the matrix variables.
\begin{equation}
	\label{dis-Hinf-ctlr}
	\begin{aligned}
		\min &  \quad t \\
		\textrm{s.t.} & \quad t>0, s>0, \\
		& \quad Y\succ 0, R\succ 0, \\
		& \quad  I_{m+n} \preceq s\begin{bmatrix} Y & 0 \\ 0 & R \end{bmatrix} \preceq tI_{m+n}, \\
		& \quad \textrm{the set of LMIs (\ref{dis-Hinf-ctlr-form}) is satisfied}.		
	\end{aligned}
\end{equation}

\subsection{Illustrative Example 2}

\textcolor{blue}{
In this section, we aim at illustrating our hierarchical decentralized robust $H_{\infty}$ controller synthesis. 
Let us consider a network of $30$ identical agents whose model is described by (\ref{agent-perturbed}) with similar system matrices $A,B$ as that in the illustrative example 1 in Section \ref{num-1}, 
and $C=[1,1]$; $D=0.3$; $B_{d}=[0,0.5]^T$. Moreover, the disturbances on the agents' models are assumed to be a white noise with magnitude $1$. 
}

\textcolor{blue}{
Consequently, we assume that the interconnection structure among agents is described by a Cartesian graph product of two chain graphs with $5$ and $6$ nodes, and choose $K$ to be a Laplacian matrix corresponding to that topology with all edge weights are equal to $1$. 
}

\begin{figure}[ht!]
		\centering
		\includegraphics[scale=0.5]{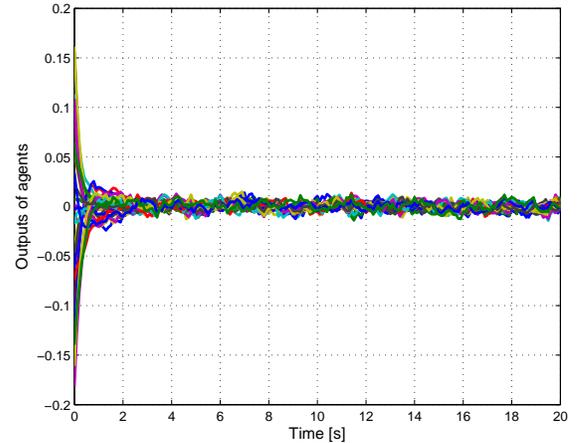}
		\caption{\textcolor{blue}{Hierarchical decentralized $H_{\infty}$ control of the given MAS in presence of a white noise with magnitude equals to $1$.} }
		\label{dis_robust_opt_Hinf}
	\end{figure}

\begin{figure}[ht!]
		\centering
		\includegraphics[scale=0.5]{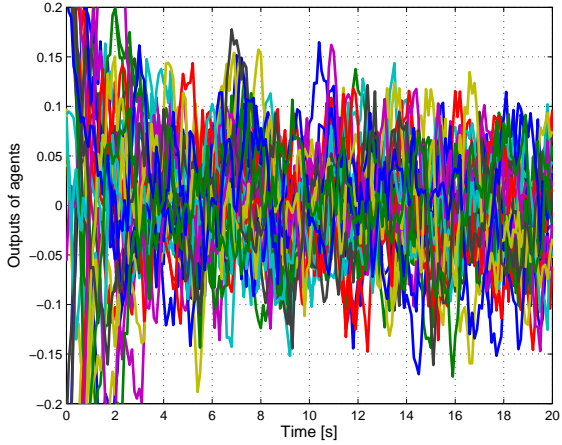}
		\caption{\textcolor{blue}{Fully decentralized $H_{\infty}$ control of the given MAS in presence of a white noise with magnitude equals to $1$.} }
		\label{dec_robust_opt_Hinf}
	\end{figure}

\textcolor{blue}{
Now, employing our proposed hierarchical decentralized robust $H_{\infty}$ controller design in Theorem \ref{Hinf-design-1}, we solve the LMI feasibility problem (\ref{Hinf-check}) with $\epsilon=1$ using MATLAB and \texttt{CVX} \cite{Boyd:2015}. Then we obtain $F_{1}=[2.9904,2.9187]$, $F_{2}=[0.0696,0.0905]$. 
The responses of agents with those controller gains are exhibited in Figure \ref{dis_robust_opt_Hinf}. 
Next, we solve a similar LMI problem with (\ref{Hinf-check}) but without $\gamma_{i}$ and $G_{2}$, i.e., a local LMI problem when the agents are not connected, and obtain $F_{1}=[1.4413,1.2911]$. 
This scenario corresponds to the fully decentralized control where there is no cooperation among agents and the agents are independently controlled by identical local controllers $F_{1}$. 
The simulation result for this fully decentralized robust $H_{\infty}$ control is shown in Figure \ref{dec_robust_opt_Hinf}. 
}


\textcolor{blue}{
Subsequently, we immediately see that the control performance provided by our proposed hierarchical decentralized robust $H_{\infty}$ controller is much better than the fully decentralized one  
in terms of both the convergence time and the magnitudes of agents' outputs. In other words, by letting agents cooperate, our hierarchical decentralized robust $H_{\infty}$ controller better eliminates the effect of disturbances to the agents' outputs and is more effective than the fully decentralized robust $H_{\infty}$ controller.  
}

\textcolor{blue}{
Finally, to validate the scalability of the proposed approach, we verify the solving time of the LMI problem (\ref{Hinf-check}) with different numbers of agents. 
For the above simulation, the solving time on a personal computer equipped with a Core i$7$ 3.4-GHz CPU is approximately $0.8$ s. 
On the other hand, if the number of agents is $300$, $600$, and $900$ agents, the solving times on the same computer are about $6.3$ s, $12.6$ s, and $19$ s, respectively, 
which seems to be linearly increased with respect to the increased number of agents. 
Thus, our proposed approach is promising even with a big number of agents.    
}

%

\section{Hierarchical Decentralized $H_{2}$ Design}
\label{H2-syn}

\subsection{General Hierarchical Decentralized $H_{2}$ Controller}

The synthesis of hierarchical decentralized robust $H_{2}$ controllers is similar to the derivation of hierarchical decentralized robust $H_{\infty}$ controllers, which searches for the existence of distributed controller gains  $F_1,F_2\in\mathbb{R}^{m\times n}$ such that the $H_{2}$ norm of the closed-loop dynamics is lower than a given bound $\epsilon>0$. The design of such controllers is presented in the following theorem. 

\begin{thm}
\label{H2-design-1}
Suppose that $K\in\mathbb{K}^{+}_{s}$ is chosen. Then the controller gains $F_1,F_2$, which let the closed-loop MAS without disturbances stabilized  and its $H_{2}$ norm less than a given $\epsilon>0$ in presence of disturbances/noise, is calculated by  (\ref{F1-F2-Hinf}) 
if there exist $Y=Y^T\in\mathbb{R}^{n\times n},Y \succ 0$, $G_1,G_2\in\mathbb{R}^{m\times n}$, and $W_i=W_i^T\in\mathbb{R}^{p\times p},W_i \succ 0$, $i=1,\ldots,N$ satisfy the following set of LMIs,
\begin{equation}
	\label{H2-check}
	\begin{aligned}
		& \begin{bmatrix}
				\mathrm{sym}(\alpha_{i}AY-\alpha_{i}B\tilde{G}_i) & B_d \\
				B_d^T & -I_{p}
			\end{bmatrix} \prec 0, \\
		& \begin{bmatrix}
				\alpha_{i}Y & \alpha_{i}(CY-D\tilde{G}_i)^T \\
				\alpha_{i}(CY-D\tilde{G}_i) & W_i
			\end{bmatrix} \succ 0, \\	
		& \sum_{i=1}^{N}\mathrm{trace}(W_i)<\epsilon^2,		
	\end{aligned}
\end{equation}		
for all $i=1,\ldots,N$, where $\tilde{G}_i=G_1+\gamma_{i}G_2$, $\gamma_{i}$ are eigenvalues of $K$, and $\alpha_{i}$ are chosen positive parameters.	
\end{thm}

\begin{IEEEproof}
See Appendix \ref{apdx-2}. 
\end{IEEEproof}

\subsection{LQR-based Hierarchical Decentralized $H_{2}$ Controller}

Like in the scenario of hierarchical decentralized $H_{\infty}$ controller design, we can here combine the $H_{2}$ design in Theorem \ref{H2-design-1} and the inverse LQR optimal design in 
Proposition \ref{inverse-LQR} into a unique LMI problem as in the following theorem. 
\begin{thm}
\label{H2-design-g}
For a given $\epsilon>0$ and a Laplacian matrix $K$, let $Y=Y^T\in\mathbb{R}^{n\times n},Y\succ 0$ and 
$R_1,R_2 \in \mathbb{R}^{m\times m},R_1=R_1^T,R_2=R_2^T,R_1,R_2\succ 0$ be the solutions of the following set of LMIs,
\begin{equation}
	\label{dis-H2-ctlr-form}
	\scalebox{0.95}{$
	\begin{aligned}
		& \begin{bmatrix}
				\mathrm{sym}(\alpha_{i}AY-\alpha_{i}B(R_1+\gamma_iR_2)) & B_d \\
				B_d^T & -I_{p}
			\end{bmatrix} \prec 0, \\
		& \begin{bmatrix}
				\alpha_{i}Y & \alpha_{i}(CY-D(R_1+\gamma_iR_2))^T \\
				\alpha_{i}(CY-D(R_1+\gamma_iR_2)) & W_i
			\end{bmatrix} \succ 0, \\	
		& \sum_{i=1}^{N}\mathrm{trace}(W_i)<\epsilon^2, \\		
		&  
			\begin{bmatrix}
				\mathrm{sym}(AY-BR_1B^T) & BR_1 \\
				R_1B^T & -R_1
			\end{bmatrix} \preceq 0,			
	\end{aligned}
	$}
\end{equation}
for all $i=1,\ldots,N$, where $\gamma_{i}$ are eigenvalues of $K$.	 
Then the distributed controller $u=-[I_{N}\otimes(R_1B^TY^{-1})+K\otimes(R_2B^TY^{-1})]$ makes the MAS (\ref{MAS-perturbed}) without disturbances stabilized and its $H_{2}$ norm in presence of disturbances less than $\epsilon$. Furthermore, this distributed controller is globally optimal with respect to the LQR performance index (\ref{p-index}) with the weighting matrices defined in 
(\ref{w-matrices-1}) and (\ref{w-matrices-2}), where $Q_1=-A^TY^{-1}-Y^{-1}A+Y^{-1}BR_{1}B^TY^{-1}$ and $Q_2=Y^{-1}BR_2B^TY^{-1}$. 
\end{thm}

\smallskip

To achieve a unique distributed $H_{2}$ consensus controller, we propose the following convex optimization problem with additional slack variables \cite{Boyd:2004} for minimizing the condition numbers of the matrix variables.
\begin{equation}
	\label{dis-H2-ctlr}
	\begin{aligned}
		\min &  \quad t \\
		\textrm{s.t.} & \quad t>0, s>0, \\
		& \quad Y\succ 0, R\succ 0, \\
		& \quad  I_{m+n} \preceq s\begin{bmatrix} Y & 0 \\ 0 & R \end{bmatrix} \preceq tI_{m+n}, \\
		& \quad \textrm{the set of LMIs (\ref{dis-H2-ctlr-form}) is satisfied}.		
	\end{aligned}
\end{equation}

\subsection{Illustrative Example 3}

\textcolor{blue}{
This example demonstrates the design of hierarchical decentralized robust $H_{2}$ controllers for the same MAS as in the illustrative example 2. Employing the design method in Section \ref{H2-syn}, 
we solve the LMI feasibility problem (\ref{H2-check}) as $\epsilon=1$ using MATLAB and \texttt{CVX} \cite{Boyd:2015} to obtain the robust $H_{2}$ controller gains $F_1=[3.2308,3.2079]$ 
and $F_2=[-0.0082,-0.0046]$. 
On the other hand, if a fully decentralized robust $H_{2}$ controller is utilized for the given MAS then $F_2=[0,0]$ and solving a local LMI problem which is similar to (\ref{H2-check}) but without 
$\gamma_{i}$ and $G_{2}$ gives us $F_1=[2.2825,2.1321]$. 
}

\begin{figure}[ht!]
		\centering
		\includegraphics[scale=0.5]{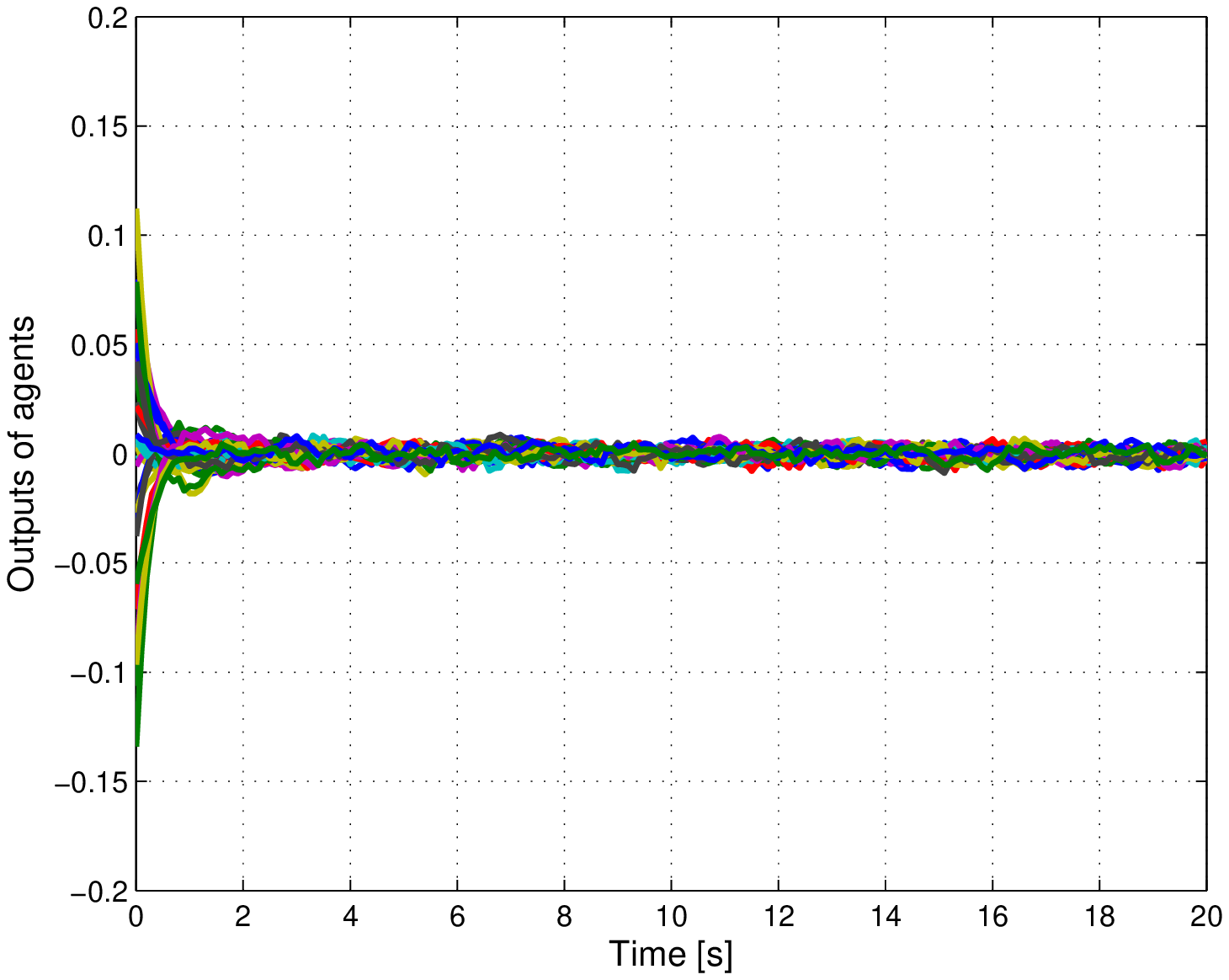}
		\caption{\textcolor{blue}{Hierarchical decentralized $H_{2}$ control of the given MAS in presence of a white noise with magnitude equals to $1$.} }
		\label{dis_robust_opt_H2}
		\includegraphics[scale=0.5]{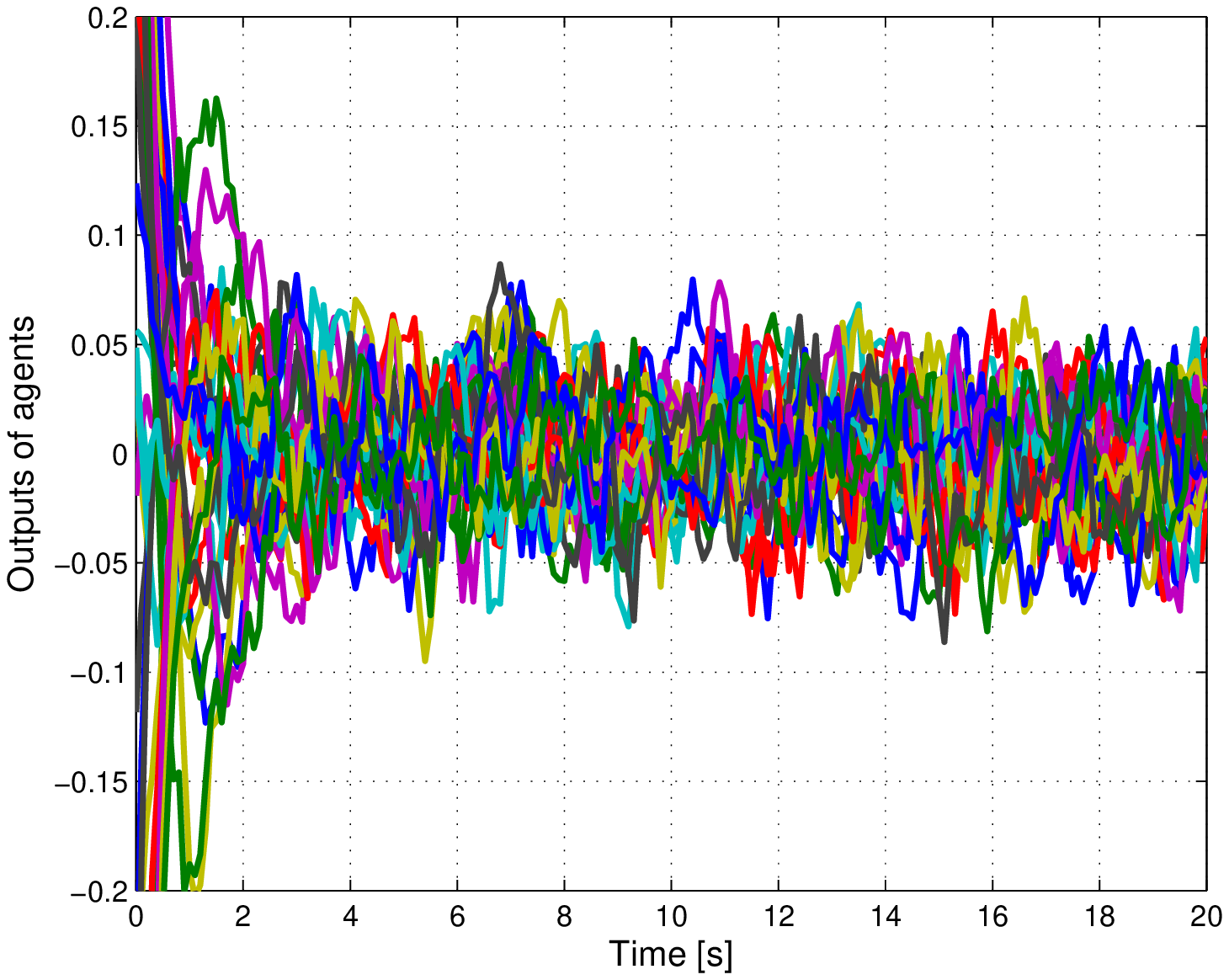}
		\caption{\textcolor{blue}{Fully decentralized $H_{2}$ control of the given MAS in presence of a white noise with magnitude equals to $1$.} }
		\label{dec_robust_opt_H2}
	\end{figure}

\textcolor{blue}{
The simulation results for the hierarchical decentralized and fully decentralized robust $H_{2}$ controllers are then displayed in Figure \ref{dis_robust_opt_H2}--\ref{dec_robust_opt_H2}. 
Obviously, the outputs of agents controlled by the hierarchical decentralized robust $H_{2}$ controller exhibit much better responses than those controlled by the fully  decentralized robust $H_{2}$ controller. More specifically, both the convergence time and the magnitudes of agents' outputs in the former case are much smaller than that in the latter case. 
This clearly shows the effectiveness of our proposed hierarchical decentralized robust $H_{2}$ controller design.   
}
	
\textcolor{blue}{
Similarly to the scenario of robust $H_{\infty}$ design, we solve the LMI problem (\ref{H2-check}) for networks of $30$, $300$, $600$, and $900$ agents and obtain small the approximated solving times  
$1.12$ s, $9.4$ s, $18.8$ s, and $28.6$ s, respectively, which are almost linearly with the increase on the number of agents. 
These show the scalability of our approach for large-scale multi-agent systems.     
}

%

\section{CONCLUSIONS}
\label{sum}

New approaches for designing hierarchical decentralized stabilizing and robust stabilizing controllers for linear homogeneous MASs have been presented in this paper. 
The remarkable features of the proposed approach are as follows. 
First, the hierarchical decentralized stabilizing controller is globally optimal w.r.t some LQR performance index. 
Moreover, its order can be reduced by suitably selecting the LQR weighting matrices.  
Second, the hierarchical decentralized robust $H_{\infty}$ and $H_{2}$ controller designs are formulated as distributed convex optimization problems with LMI constraints, 
of which their conservatism is reduced by introducing some scaled parameters resulting in non-common Lyapunov variables. 
\textcolor{blue}{Lastly, the effectiveness and scalability for large-scale MASs of the proposed approaches are clearly illustrated through several numerical examples.  }

An open problem for the next researches is to derive necessary and sufficient conditions for hierarchical decentralized robust $H_{\infty}$ and $H_{2}$ controller designs to completely remove the conservatism of current results.


%

\appendices

\section{Proof of Theorem \ref{Hinf-design-1}}
\label{apdx-1}

It can be deduced from (\ref{distributed-norms}) that $\|G_{\tilde{y}\tilde{d}}(s)\|_{\infty} < \epsilon$ if and only if $\|G_{\tilde{y}_i\tilde{d}_i}(s)\|_{\infty} < \epsilon \;\forall\; i=1,\ldots,N$. 
Let $$\tilde{A}_{i}=A-B(F_1+\gamma_{i}F_2),\tilde{C}_{i}=C-D(F_1+\gamma_{i}F_2),$$ 
then based on the result of $H_{\infty}$ controller design for single LTI systems \cite{Scherer:2005}, the decomposed system (\ref{MAS-perturbed-2}) has  
$\|G_{\tilde{y}_i\tilde{d}_i}(s)\|_{\infty} < \epsilon$ if and only if there exist $F_1,F_2\in\mathbb{R}^{m\times n}$ and $Y_{i}=Y_{i}^T\in\mathbb{R}^{n\times n},Y_{i} \succ 0,i=1,\ldots,N$ 
satisfy the following set of matrix inequalities,
\begin{equation}
	\label{Hinf-check-1}
	\begin{bmatrix}
		\mathrm{sym}(\tilde{A}_iY_{i})+\frac{1}{\epsilon^2}B_dB_d^T & (\tilde{C}_iY_{i})^T \\
		\tilde{C}_iY_{i} & -I_{p} \\
	\end{bmatrix} \prec 0.  
\end{equation}
Let $G_{1,i}=F_1Y_{i}$ and $G_{2,i}=F_2Y_{i}$ for all $i=1,\ldots,N$ then (\ref{Hinf-check-1}) become LMIs, which is solvable. Nevertheless, once we obtain $G_{1,i},G_{2,i}$ and $Y_{i}$ by solving 
those LMIs, generally it cannot be guaranteed that $G_{1,i}Y_{i}^{-1}$ with $i=1,\ldots,N$ are all equal. The same situation occurs for $G_{2,i}Y_{i}^{-1},i=1,\ldots,N$.  
One way to overcome that situation is to set $Y_{i}=\alpha_{i}Y$ and $G_{1,i}=\alpha_{i}G_{1},G_{2,i}=\alpha_{i}G_{2}$ where $G_{1} \triangleq F_1Y,G_{2} \triangleq F_2Y$, $\alpha_{i}>0$, $i=1,\ldots,N$, 
and $Y=Y^T\in\mathbb{R}^{n\times n},Y \succ 0$,   
which makes (\ref{Hinf-check-1}) be equivalent to (\ref{Hinf-check}) and $F_1,F_2$ can be found as in (\ref{F1-F2-Hinf}).


\section{Proof of Theorem \ref{H2-design-1}}
\label{apdx-2}

The formulation (\ref{distributed-norms}) shows that $\|G_{\tilde{y}\tilde{d}}(s)\|_{2} < \epsilon$ if and only if sum of square of $H_{2}$ norms of decomposed systems (\ref{MAS-perturbed-2}) is less than $\epsilon^2$. Hence, based on the synthesis of $H_{2}$ controllers for single LTI systems \cite{Scherer:2005}, the design of a distributed $H_{2}$ controller for the given MAS (\ref{MAS-perturbed}) can be formulated as a feasibility problem with the following set of matrix inequalities,
\begin{equation}
	\label{H2-check-1}
	\begin{aligned}
		& \begin{bmatrix}
				\mathrm{sym}(\tilde{A}_iY_{i}) & B_d \\
				B_d^T & -I_{p}
			\end{bmatrix} \prec 0, \\
		& \begin{bmatrix}
				Y_{i} & (\tilde{C}_iY_{i})^T \\
				\tilde{C}_iY_{i} & W_i
			\end{bmatrix} \succ 0, \\	
		& \sum_{i=1}^{N}\mathrm{trace}(W_i)<\epsilon^2,		
	\end{aligned}
\end{equation}
where $F_1,F_2\in\mathbb{R}^{m\times n}$, $Y_{i}=Y_{i}^T\in\mathbb{R}^{n\times n},Y_{i} \succ 0,i=1,\ldots,N$, and $W_{i}=W_{i}^T\in\mathbb{R}^{n\times n},W_{i} \succ 0,i=1,\ldots,N$. 
Similarly to the case of distributed $H_{\infty}$ controller design in Appendix \ref{apdx-1}, it is reasonable to set $Y_{i}=\alpha_{i}Y$ and 
$G_{1,i}=\alpha_{i}G_{1},G_{2,i}=\alpha_{i}G_{2}$ where $G_{1} \triangleq F_1Y,G_{2} \triangleq F_2Y$, $\alpha_{i}>0$, $i=1,\ldots,N$, 
and $Y=Y^T\in\mathbb{R}^{n\times n},Y \succ 0$. 
Obviously, this makes (\ref{H2-check-1}) be equivalent to (\ref{H2-check}).

\section*{Acknowledgment}

The authors would like to thank the associate editor and the anonymous reviewers for their valuable comments and suggestions that significantly help improving the paper. 

This research is partially supported by Hitech Research Center, projects for private universities, supplied from the Ministry of Education, Culture, Sports, Science and Technology, Japan.

\ifCLASSOPTIONcaptionsoff
  \newpage
\fi



%

\bibliographystyle{IEEEtran}

\bibliography{References}

%
%

%

%
%
%
%
%
%
%
%
%
\end{document}